\documentclass[12pt,epsf]{article}
\usepackage{graphicx,amsmath,amssymb}
\begin{document}
%%%%%%%%%%%%%%%%%%%%%%%%%%%%%%%%%%%%%%%%%%%

\def\a{\alpha}
\def\b{\beta}
\def\c{\varepsilon}
\def\d{\delta}
\def\e{\epsilon}
\def\f{\phi}
\def\g{\gamma}
\def\h{\theta}
\def\k{\kappa}
\def\l{\lambda}
\def\m{\mu}
\def\n{\nu}
\def\p{\psi}
\def\q{\partial}
\def\r{\rho}
\def\s{\sigma}
\def\t{\tau}
\def\u{\upsilon}
\def\v{\varphi}
\def\w{\omega}
\def\x{\lambda}
\def\y{\eta}
\def\z{\zeta}
\def\D{\Delta}
\def\G{\Gamma}
\def\H{\Theta}
\def\L{\Lambda}
\def\F{\Phi}
\def\P{\Psi}
\def\S{\Sigma}

\def\o{\over}
\def\beq{\begin{eqnarray}}
\def\eeq{\end{eqnarray}}
\newcommand{\gsim}{ \mathop{}_{\textstyle \sim}^{\textstyle >} }
\newcommand{\lsim}{ \mathop{}_{\textstyle \sim}^{\textstyle <} }
\newcommand{\vev}[1]{ \left\langle {#1} \right\rangle }
\newcommand{\bra}[1]{ \langle {#1} | }
\newcommand{\ket}[1]{ | {#1} \rangle }
\newcommand{\EV}{ {\rm eV} }
\newcommand{\KEV}{ {\rm keV} }
\newcommand{\MEV}{ {\rm MeV} }
\newcommand{\GEV}{ {\rm GeV} }
\newcommand{\TEV}{ {\rm TeV} }
\def\diag{\mathop{\rm diag}\nolimits}
\def\Spin{\mathop{\rm Spin}}
\def\SO{\mathop{\rm SO}}
\def\O{\mathop{\rm O}}
\def\SU{\mathop{\rm SU}}
\def\U{\mathop{\rm U}}
\def\Sp{\mathop{\rm Sp}}
\def\SL{\mathop{\rm SL}}
\def\tr{\mathop{\rm tr}}

\def\IJMP{Int.~J.~Mod.~Phys. }
\def\MPL{Mod.~Phys.~Lett. }
\def\NP{Nucl.~Phys. }
\def\PL{Phys.~Lett. }
\def\PR{Phys.~Rev. }
\def\PRL{Phys.~Rev.~Lett. }
\def\PTP{Prog.~Theor.~Phys. }
\def\ZP{Z.~Phys. }

%%%%%%% added by Fumi %%%%%%%%%%
% FROM HERE
%\newcommand{\beq}{\begin{equation}}   
%\newcommand{\eeq}{\end{equation}}
\newcommand{\bea}{\begin{eqnarray}}   
\newcommand{\eea}{\end{eqnarray}}
\newcommand{\bear}{\begin{array}}  
\newcommand {\eear}{\end{array}}
\newcommand{\bef}{\begin{figure}}  
\newcommand {\eef}{\end{figure}}
\newcommand{\bec}{\begin{center}}  
\newcommand {\eec}{\end{center}}
\newcommand{\non}{\nonumber}  
\newcommand {\eqn}[1]{\beq {#1}\eeq}
\newcommand{\la}{\left\langle}  
\newcommand{\ra}{\right\rangle}
\newcommand{\ds}{\displaystyle}
\def\SEC#1{Sec.~\ref{#1}}
\def\FIG#1{Fig.~\ref{#1}}
\def\EQ#1{Eq.~(\ref{#1})}
\def\EQS#1{Eqs.~(\ref{#1})}
\def\GEV#1{10^{#1}{\rm\,GeV}}
\def\MEV#1{10^{#1}{\rm\,MeV}}
\def\KEV#1{10^{#1}{\rm\,keV}}
\def\lrf#1#2{ \left(\frac{#1}{#2}\right)}
\def\lrfp#1#2#3{ \left(\frac{#1}{#2} \right)^{#3}}
% UNTIL HERE

%%%%%%%%%%%%%%%%%%%%%%%%%%%%%%%%%%%%%%%%%%%%%%%%%%%%%%%%%%%%%%%%%%%%

\baselineskip 0.7cm

\begin{titlepage}

\begin{flushright}
UT-14-34\\
TU-977\\
IPMU14-0223\\
\end{flushright}

\vskip 1.35cm
\begin{center}
{\large \bf 
Polynomial Chaotic Inflation in Supergravity Revisited
}
\vskip 1.2cm
Kazunori Nakayama$^{a,c}$,
Fuminobu Takahashi$^{b,c}$
and 
Tsutomu T. Yanagida$^{c}$

\vskip 0.4cm

{\it $^a$Department of Physics, University of Tokyo, Tokyo 113-0033, Japan}\\
{\it $^b$Department of Physics, Tohoku University, Sendai 980-8578, Japan}\\
{\it $^c$Kavli IPMU, TODIAS, University of Tokyo, Kashiwa 277-8583, Japan}

\vskip 1.5cm

\abstract{

We revisit a polynomial chaotic inflation model in supergravity which we proposed soon after the 
Planck first data release. Recently some issues have been raised in Ref.~\cite{Linde:2014nna}, concerning the validity of our polynomial chaotic inflation model. We study the inflaton dynamics in detail, and confirm that the inflaton potential
is very well approximated by a polynomial potential for the parameters of our interest in any practical sense, and
in particular,  the spectral index and the tensor-to-scalar ratio can be estimated by single-field approximation.
This justifies our analysis of the polynomial chaotic inflation in supergravity.

}
\end{center}
\end{titlepage}

\setcounter{page}{2}

%%%%%%%%%%%%%%%%%%%%%%%%%%%%%%%%%%%%%
%\section{Introduction}
%%%%%%%%%%%%%%%%%%%%%%%%%%%%%%%%%%%%%

Inflation provides elegant solutions to several theoretical problems of the standard 
big bang cosmology such as the horizon and flatness problems \cite{Guth:1980zm,Kazanas:1980tx},
and the slow-roll inflation paradigm~\cite{Linde:1981mu, Albrecht:1982wi} 
successfully explains observations of cosmic microwave background  (CMB)
and large-scale structure.
Particularly interesting is the so called large-field inflation that can generate
a sizable tensor-to-scalar ratio, $r$, within  the reach of the on-going  and planned CMB 
experiments.  Among various large-field inflation models, the simplest one is the quadratic chaotic inflation model
proposed by Linde long time ago~\cite{Linde:1983gd}.

The Planck satellite observed the CMB temperature and polarization anisotropy 
with unprecedented accuracy. Planck released the first data with a series of papers in March 2013~\cite{Ade:2013rta}, providing tight constraints on the scalar spectral index $n_s$ and the tensor-to-scalar ratio $r$.
Soon after the Planck first data release, 
we proposed a polynomial chaotic  inflation in supergravity (SUGRA)~\cite{Nakayama:2013jka}
as an extension of Refs.~\cite{Kawasaki:2000yn,Kallosh:2010ug}.
We showed in Ref.~\cite{Nakayama:2013jka}  that the predicted values of $n_s$  and $r$ can cover almost 
entire region allowed by 
the Planck data. The inflaton dynamics was further studied in  a more general set-up in Ref.~\cite{Nakayama:2013txa}.
The polynomial chaotic inflation has gained momentum recently, especially after the BICEP2 
collaboration claimed a detection of primordial B-mode polarization~\cite{Ade:2014xna}. The dynamics of 
the polynomial chaotic inflation and its variation have been studied in Refs.~\cite{Linde:2014nna,Kobayashi:2014jga,Kallosh:2014xwa,Nakayama:2014hga}. 

In Ref.~\cite{Linde:2014nna}, several issues were raised concerning our inflation model: 1) the real component 
of the inflaton field acquires a non-zero vacuum
expectation value which depends on the inflaton field, and so, we will have no longer the simple single-field inflation;
2) the potential is not exactly polynomial as a result of 1); 3) the kinetic term of the fields will be 
non-canonical and non-diagonal; 4) there is an extra minimum.  
The purpose of the present letter is to study our polynomial chaotic inflation model in detail, 
in answer to the above issues. 

Our short answer is that, in any practical sense,  the inflaton potential
is very well approximated by a polynomial potential for the parameters of our interest, and
in particular,  the spectral index and the tensor-to-scalar ratio can be estimated by single-field approximation.
The kinetic terms can be easily diagonalized and canonically normalized by only slightly rotating the
field basis. There is an extra minimum which however is located outside of the validity region of our inflation model, and it does not affect the inflaton dynamics significantly.
The typical change of the field basis is so small that the predicted 
values of $(n_s, r)$ remain almost unchanged. 
This justifies our analysis of the polynomial chaotic inflation in SUGRA in Ref.~\cite{Nakayama:2013jka}.
Therefore, our model is a concrete realization of the polynomial chaotic inflation in SUGRA\footnote{
It was already mentioned by Linde in Ref.~\cite{Linde:2014nna} that the above issues may not be a big problem
in our polynomial chaotic inflation model~\cite{Nakayama:2013jka}.
}, for which the predicted values of the spectral index and the tensor-to-scalar ratio can cover 
almost entire region allowed by Planck.
\\

The central issue in building successful chaotic inflation models in SUGRA
is how to have a good control of the inflaton potential over super-Planckian field ranges.
A simple prescription was given in the paper~\cite{Kawasaki:2000yn}, where they 
introduce a shift symmetry on the inflaton field $\phi$ along its imaginary component:
\beq
\label{ss}
\phi \to \phi + i C,
\eeq
where $C$ is a real transformation parameter. 
The K\"ahler potential takes the following form~\cite{Kawasaki:2000yn}
\begin{equation}
	K = \frac{1}{2}(\phi+\phi^\dagger)^2 + |X|^2 + \cdots
\end{equation}
which satisfies the shift symmetry, whereas
it is explicitly broken by the superpotential of the form,
\begin{equation}
	W = mX \phi,
\end{equation}
where $X$ is a singlet chiral superfield with $R$-charge $2$. The introduction of $X$ is crucial for avoiding a negative inflaton 
potential at large field values of $\phi$. The scalar potential in SUGRA is given by
\begin{equation}
	V = e^K \left[ K^{i\bar j}(D_iW)(D_{\bar j}\bar W) - 3|W|^2 \right],
\end{equation}
where $D_iW \equiv \partial_i W + K_i W$ and $K^{i\bar j} = K_{i\bar j}^{-1}$.
Here and in what follows, we adopt the Planck units in which $M_P\simeq 2.4\times 10^{18}$\,GeV is set to be unity.
The inflaton potential is generated by the small shift symmetry breaking 
superpotential, and it is given by
\begin{equation}
	V = \frac{1}{2}m^2\varphi^2,
\end{equation}
where  $\varphi\equiv \sqrt{2}{\rm Im}(\phi)$, and both $X$ and ${\rm Re}[\phi]$ are stabilized
at the origin.
The approximate shift symmetry ensures the flatness of the potential along the imaginary 
component ${\rm Im}(\phi)$ beyond the Planck scale.

Now we move on to the polynomial chaotic inflation model~\cite{Nakayama:2013jka,Nakayama:2013txa}.
We consider the following K\"ahler potential satisfying the shift symmetry (\ref{ss}),
\begin{equation}
	K = \frac{1}{2}(\phi + \phi^\dagger)^2 + |X|^2 - c_\phi |X|^2(\phi + \phi^\dagger)^2 - c_X|X|^4 + \cdots
\end{equation}
where $c_\phi$ and $c_X$ are constants of order unity, the dots represent higher
order terms, and  a linear term of $\phi + \phi^\dag$ is dropped,  since it does not affect the inflaton 
dynamics~\cite{Nakayama:2013txa}. We introduce shift symmetry breaking terms in the superpotential 
as~\footnote{In Ref.~\cite{Nakayama:2014hga}, we proposed an extension to include multiple $X$ fields.
For instance we can consider $W = X_1 f(\phi) + X_2 g(\phi)$ to induce the scalar potential 
$V = e^K \left(|f(\phi)|^2 + |g(\phi)|^2 \right)$. By taking e.g. $f(\phi) \propto \phi$ and $g(\phi) \propto \phi^2$,
one can realize a polynomial chaotic inflation in supergravity without a cross term.
}
\bea
W = m X (\phi + k_2 \phi^2 + k_3 \phi^3 + \cdots),
\label{Wmany}
\eea
where $k_i$  represents the numerical coefficient of higher order terms.
See Ref.~\cite{Harigaya:2014qza} for the case with shift symmetry breaking terms in the K\"ahler potential.
To be  concrete, we focus on the case where the first two terms in the superpotential 
make the dominant contribution to the inflaton dynamics:
\begin{equation}
	W = m X \left(\phi + \lambda e^{i \theta} \phi^2 \right),  \label{W}
\end{equation}
where we have defined $\lambda \equiv |k_2|$ and $\theta \equiv {\rm arg}[k_2]$, and
we assume $\lambda = {\cal O}(0.1)$~\cite{Nakayama:2013jka,Nakayama:2013txa}\footnote{The cut-off scale one order of magnitude larger 
than the Planck scale can be
 understood as follows. 
Suppose that the shift symmetry is broken by various Planck-suppressed shift symmetry breaking terms.
Then, if the kinetic term coefficient happens to be enhanced by by a factor of ${\cal O}(10-100)$, 
all the higher order terms are suppressed when they are expressed in terms of the canonically normalized field.
Such an enhancement may be realized if there are many singlet scalars whose kinetic term coefficients are 
subject to a certain random distribution~\cite{Nakayama:2014hga}. }.  See also Appendix of this letter
for another case. 

First let us see that $X$ is stabilized at the origin $X = 0$ during inflation.
This is because $X$ obtains an inflaton-dependent mass term as
\begin{equation}
	V \supset 4c_X |X|^2 m^2 |\phi +  \lambda e^{i \theta} \phi^2|^2 \simeq 12c_X H^2 |X|^2,
\end{equation}
where $H$ denotes the Hubble parameter during inflation.
Therefore, for $c_X \gtrsim \mathcal O(0.1)$, $X$ obtains a mass of order of the Hubble scale
and it is stabilized at the origin during inflation.\footnote{
	By including the constant term in the superpotential $W_0 = m_{3/2}$, where $m_{3/2}$ denotes the gravitino mass,
	the minimum of $X$ during inflation slightly is deviated from $X=0$. Such a shift is safely neglected
	as long as $m_{3/2}$ is much smaller than the inflaton mass.
}
Thus in the following analysis we take $X=0$.
Note that the inflaton field is canonically normalized for $X=0$.

%%%%%%%%%%%%%%%%
\begin{figure}
\begin{center}
\includegraphics[scale=1.5]{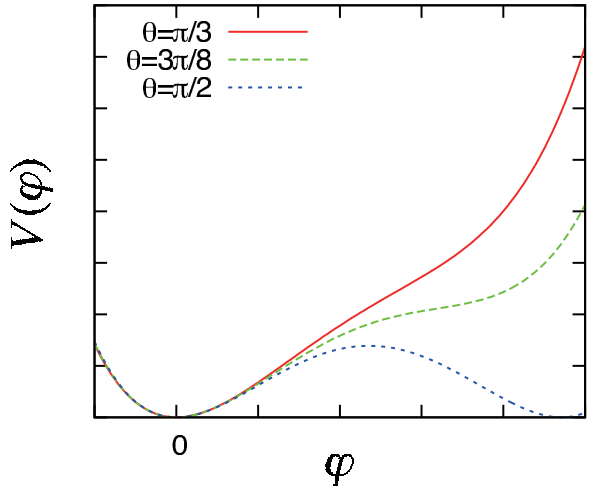}
\caption{ 
	Schematic picture for the scalar potential (\ref{pol}).
	}
\label{fig:pot12}
\end{center}
\end{figure}
%%%%%%%%%%%%%%%%

Let us decompose the scalar field $\phi$ as
\begin{equation}
	\phi = \frac{1}{\sqrt{2}}(\chi + i\varphi),
\end{equation}
where $\chi$ and $\varphi$ are real and imaginary components, respectively. 
As noted above, $\varphi$ can develop a field value much larger than the Planck scale because of the shift symmetry.
On the other hand, $\chi$ obtains a Hubble-induced mass and stabilized at sub-Planckian field values $|\chi| \ll 1$.
In Refs.~\cite{Nakayama:2013jka,Nakayama:2013txa} we approximated $\chi \approx 0$ and obtained the inflaton potential
as
\begin{equation}
V \;\simeq\; V_{\rm inf}(\varphi) = \frac{1}{2}m^2 \varphi^2
	\left( 1 - \sqrt{2}  \lambda \sin \theta\, \varphi + \frac{\lambda^2}{2} \varphi^2 \right).
		  \label{pol}
\end{equation}
The potential shape is shown in Fig.~\ref{fig:pot12}. 
One can see that the inflaton potential changes its form as one varies $\theta$.  Therefore, as long as the approximation is
valid,  the polynomial chaotic inflation can be realized by the first two terms in (\ref{Wmany}).

In Ref.~\cite{Kallosh:2010ug} the superpotential  was extended to be the form
of $W = X f(\phi)$, where $f(\phi)$ is an arbitrary holomorphic function, and it was shown that the real component
of $\phi$ can be stabilized at the origin for a certain class of $f(\phi)$, where the coefficients are either purely 
real or imaginary depending on the definition of the shift symmetry.  In this case, 
the polynomial chaotic inflation can be realized for a certain combination of three terms 
 in the superpotential~\cite{Kallosh:2014xwa}.

The above inflaton potential (\ref{pol}) is slightly modified once one takes account of the fact that
the real component $\chi$ acquires a vacuum expectation value which depends on $\varphi$,
as pointed out in Ref.~\cite{Linde:2014nna}. 
To see this, we expand the full SUGRA potential in $\chi$.
Then we obtain
\bea
	V&=&V_{\rm inf}(\varphi) \left(1+(2c_\phi+1)\chi^2\right)+ \frac{1}{\sqrt{2}}\lambda\cos\theta m^2\varphi^2\chi 
	+  \frac{1}{2} \lambda^2  m^2\varphi^2\chi^2+ \cdots.
\label{Vhigher}
\eea
%%
%for $\varphi\gg 1$.
Thus $\chi$ obtains a mass of order Hubble scale and it is stabilized at 
\bea
\chi_{\rm min} \;\approx\; - \frac{\sqrt{2} \lambda \cos\theta}{2 c_\phi+1} \frac{1}{1-\sqrt{2} \lambda \sin \theta \varphi + \frac{\lambda^2}{2} \varphi^2}
\label{vsugra}
\eea
during inflation. It is the $\varphi$-dependence of $\chi_{\rm min}$ that modifies the inflationary path and the inflaton
potential, because the constant part of $\chi_{\rm min}$ simply modifies the coefficients of the polynomial potentials
as can be seen from (\ref{Vhigher}).
In fact,
it is easy to see that the $\varphi$-dependence of $\chi_{\rm min}$ is rather suppressed:
\bea
\frac{\partial \chi_{\rm min}}{\partial \varphi} = {\cal O}(\lambda^2),
\eea
which becomes even smaller for $\varphi \gtrsim \lambda^{-1}$.
For $\lambda = {\cal O}(0.1)$,  therefore, the modification of the inflationary path as well as to the inflaton potential
is at most of order ${\cal O}(1)$\% level. The corrections to $n_s$ and $r$ are expected to be of a similar 
order.

\begin{figure}
\begin{center}
\includegraphics[scale=1.2]{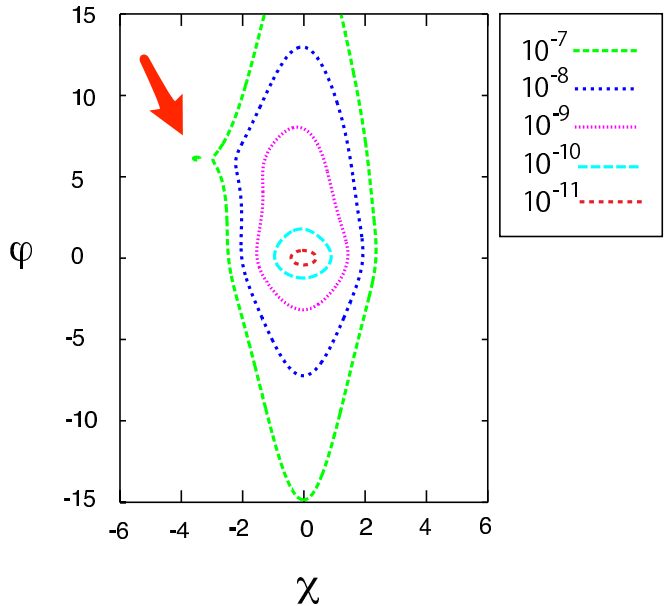}
\caption{ 
	Contours of the scalar potential in Planck unit for the model (\ref{W}), $\theta=\pi/3$ and $\lambda= 0.2$. 
	We have taken $m=10^{-5}$ and $c_\phi = 0$. The inflaton path is well approximated by $\chi \simeq 0$. The arrow shows the second minimum in (\ref{min}), which is located outside the validity region of our model. See the text for details.
}
\label{fig:pot}
\end{center}
\end{figure}

The contour of the inflaton potential in the complex $\phi$ plane is shown in Fig.~\ref{fig:pot}.
It is seen that there are two global minima at
\begin{equation}
	\phi=0,~~~\phi=-\lambda^{-1} e^{-i\theta},
	\label{min}
\end{equation}
and the potential is deformed asymmetrically. Note that the second
minimum is super-Planckian along the real component for $\lambda = {\cal O}(0.1)$ and  a general value of $\theta$,
and therefore it is outside the validity region of our inflation model. We draw the contour
only for visualization purpose, simply by extrapolating the SUGRA potential to $|\chi| \gg 1$. 
In any case, there is an exponential potential barrier between these two minima, and the effect of deformation is not significant. 
Also for $c_\phi > 0$, $\chi$ becomes heavier and the inflationary trajectory becomes  closer to $\chi = 0$.
For the reasons stated above, we expect that we can approximately set $\chi \simeq 0$ during inflation 
as in our previous study~\cite{Nakayama:2013jka,Nakayama:2013txa}.  Next we study the inflaton dynamics
numerically to show this explicitly.

%%%%%%%%%%%%%%%%
%%%%%%%%%%%%%%%%

In order to see how large is the effect of the deformation of the inflaton potential on the predicted values of $n_s$ and $r$, 
we have performed numerical calculation  using the full SUGRA potential.
We have solved the two field inflaton dynamics $\chi$ and $\varphi$ and identify the inflaton direction $\tilde\varphi$ as a mixture of
$\chi$ and $\varphi$ as $\tilde\varphi = c \varphi + s \chi$ where~\cite{Gordon:2000hv}
\begin{equation}
	c = \frac{V_\varphi}{\sqrt{V_\varphi^2+V_\chi^2}},~~~s=\frac{V_\chi}{\sqrt{V_\varphi^2+V_\chi^2}}.
\end{equation}
with subscript $\varphi$ and $\chi$ being the derivative with respect to it.
The scalar spectral index and the tensor-to-scalar ratio is obtained from $n_s = 1-6\epsilon + 2\eta$
and $r=16\epsilon$ where 
\begin{equation}
	\epsilon = \frac{1}{2}\left(\frac{V'}{V}\right)^2,~~\eta = \frac{V''}{V},
\end{equation}
with prime denoting the derivative with respect to $\tilde\varphi$:
they are given by $V' = cV_\varphi + sV_\chi$, $V'' = c^2 V_{\varphi\varphi} + 2sc V_{\varphi\chi} + s^2 V_{\chi\chi}$.
They are evaluated at the point where the e-folding number is $N_e$.
In the numerical analysis, we take $N_e = 60$.

%%%%%%%%%%%%%%%%
\begin{figure}[t!]
\begin{center}
\includegraphics[scale=1.8]{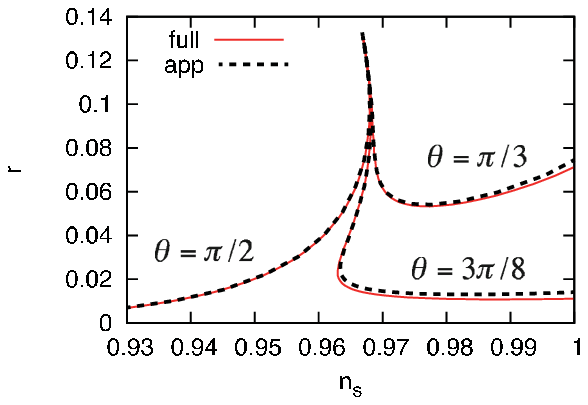}
\vskip 1cm
\includegraphics[scale=1.8]{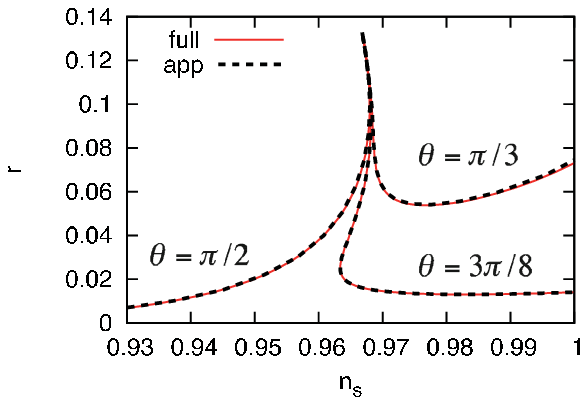}
\caption{ 
	Comparison of $(n_s, r)$ between full SUGRA result and approximate result for $\theta=\pi/3, 3\pi/8$ and $\pi/2$
	for the model (\ref{W}).
	We have taken $c_\phi=0$ (upper panel) and $c_\phi=1$ (lower panel).
}
\label{fig:ns}
\end{center}
\end{figure}
%%%%%%%%%%%%%%%%

The result is shown by (red) solid lines in Fig.~\ref{fig:ns} for $\theta=\pi/3, 3\pi/8, \pi/2$ 
and $c_\phi=0$ (upper panel) and $c_\phi=1$ (lower panel). We have varied $\lambda$ in the range of 
$\lambda=0 \sim 0.2$. The Planck normalization on the density perturbation is imposed.
For comparison, we have also plotted the result for the approximate case where $\chi$ is set to be zero ((black) dashed lines).
For $c_\phi=0$ (upper panel), the results based on the full SUGRA potential agree well with the one based
on the single-field approximation. 
The discrepancy between these two results are actually small:
the change of $(n_s,r)$ can be absorbed by small change of $\theta$.
This is because the $\varphi$-dependence of $\chi_{\rm min}$  stretches the inflaton potential by a small amount, which effectively  amounts to shifting the parameters $\lambda$ and $\theta$ slightly.
The lower panel is the same plot but for $c_\phi=1$.
As can be clearly seen, the full SUGRA results almost coincide with those of polynomial potential (\ref{pol}).
In any case, even if we take account of the full SUGRA potential,  the predicted values of $(n_s, r)$ of our model 
cover the almost entire region allowed by Planck, as in the case where the inflaton potential is approximated by a polynomial.
Considering the uncertainties on $N_e$ and observational errors of $n_s$ and $r$, the single-field approximation 
with a polynomial potential is sufficient to estimate the prediction of $n_s$ and $r$. We have also estimated $n_s$ and $r$ based on $\delta N$-formalism solving the two field dynamics numerically, and obtained consistent results.

%%%%%%%%%%%%%%%%%%%%%%%%%%%%%%%%%%%%%%%%%%%%
\section*{Acknowledgments}
%%%%%%%%%%%%%%%%%%%%%%%%%%%%%%%%%%%%%%%%%%%%

We thank Andrei Linde for pointing out the issues of the polynomial chaotic inflation model.
This work was supported by the Grant-in-Aid for Scientific Research on
Innovative Areas (No.23104008 [FT], No.26104009 [TTY]),  Scientific Research (A) (No.26247042 [KN, FT]),
Scientific Research (B) (No.26287039 [FT, TTY]),  
Young Scientists (B) (No.26800121 [KN], No.24740135) [FT]),  and Inoue Foundation for Science [FT].  
This work was also supported by World Premier International Center Initiative (WPI Program), MEXT, Japan.

\appendix
\section{The case of $W = X(\phi + \phi^3)$}
\begin{figure}
\begin{center}
\includegraphics[scale=1.2]{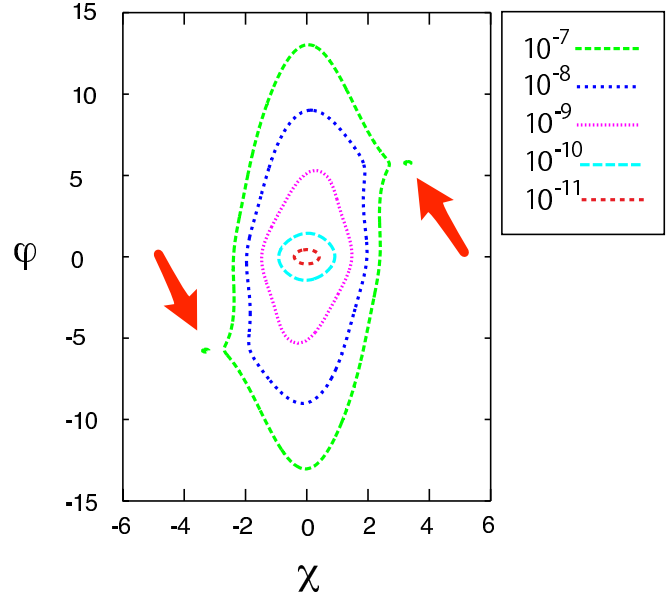}
\caption{ 
	Contours of the scalar potential in Planck unit for the model (\ref{W2}), $\theta=\pi/3$ and $\xi= 0.045$. 
	We have taken $m=10^{-5}$ and $c_\phi = 0$.
}
\label{fig:pot2}
\end{center}
\end{figure}

\begin{figure}
\begin{center}
\includegraphics[scale=1.5]{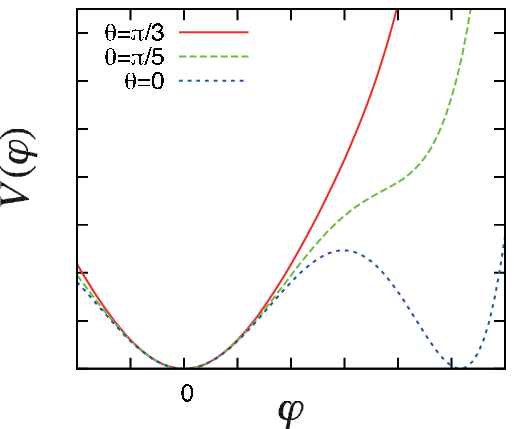}
\caption{ 
	Schematic picture for the scalar potential (\ref{Vapp2}).
	}
\label{fig:pot13}
\end{center}
\end{figure}

\begin{figure}
\begin{center}
\includegraphics[scale=1.8]{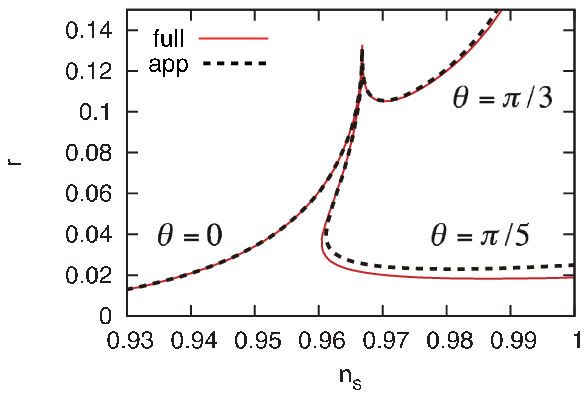}
\vskip 1cm
\includegraphics[scale=1.8]{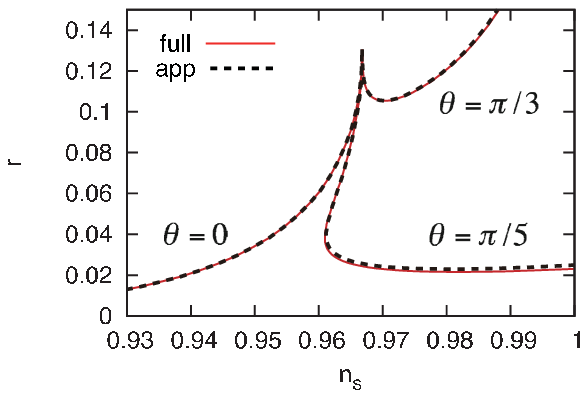}
\caption{ 
	Comparison of $(n_s, r)$ between full SUGRA result and approximate result for $\theta=0, \pi/5$ and $\pi/3$
	for the model (\ref{W2}). We have taken $c_\phi=0$ (upper panel) and $c_\phi=1$ (lower panel).
}
\label{fig:ns2}
\end{center}
\end{figure}

In this Appendix we similarly study the case of different choice of the superpotential:
\bea
W = mX(\phi + \xi e^{i\theta}\phi^3).  \label{W2}
\eea
where $\xi$ is a real constant.
This form of the superpotential is of particular interest because it is ensured by imposing a $Z_2$ symmetry on $\phi$ and $X$
and, because of the $Z_2$ symmetry, this model is free from the gravitino overproduction from the inflaton 
decay~\cite{Kawasaki:2006gs,Asaka:2006bv,Endo:2006qk,Endo:2007ih,Endo:2007sz}.
The contours of the scalar potential is shown in Fig.~\ref{fig:pot2} for $\theta=\pi/3$, $\xi= 0.045$ and $c_\phi=0$.
It is seen that the potential is slightly deformed in the direction of $\chi$ and there exist three golobal minima:
\begin{equation}
	\phi=0,~~~\phi^2=-\xi^{-1}e^{-i\theta}.
\end{equation}
We obtain the following approximate inflaton potential by setting $\chi \simeq 0$:
\begin{equation}
V \;\simeq\; V_{\rm inf}(\varphi) = \frac{1}{2}m^2 \varphi^2
	\left( 1 - \xi \cos \theta\, \varphi^2 + \frac{\xi^2}{4} \varphi^4 \right).  \label{Vapp2}
\end{equation}
A schematic picture for the scalar potential (\ref{Vapp2}) is shown in Fig.~\ref{fig:pot13}.
We focus on the case of $\xi = {\cal O}(0.01)$ where the second term affects the inflaton dynamics during the last $50-60$ e-foldings. 
We have numerically solved the inflaton dynamics under the full SUGRA potential
 and the result is plotted in Fig.~\ref{fig:ns2} for $\theta=0, \pi/5$ and $\pi/3$. 
We have taken $c_\phi=0$ (upper panel) and $c_\phi=1$ (lower panel).
Similarly to the case studied in the main text, 
it is seen that the difference between approximate results and the full results are very small.

%%%%%%%%%%%%%%%%
%%%%%%%%%%%%%%%%

%%%%%%%%%%%%%%%%
%%%%%%%%%%%%%%%%

%%%%%%%%%%%%%%%%
%%%%%%%%%%%%%%%%

%%%%%%%%%%%%%%%%%%%%%%%%%%%%%%%%%%%%%

%%%%%%%%%%%%%%%%%%%%%%%%%%%%%%%%%%%%%

\end{document}